\documentclass[twoside]{article}

\usepackage[accepted]{aistats2025}
\usepackage[round]{natbib}

\usepackage{times}
\usepackage{latexsym}
\usepackage[T1]{fontenc}
\usepackage[utf8]{inputenc}
\usepackage{microtype}
\usepackage{inconsolata}
\usepackage{graphicx}
\usepackage{caption}
\usepackage{amsmath}
\usepackage{amssymb}
\usepackage{comment}
\usepackage{booktabs}
\usepackage{hyperref}
\usepackage[overload]{empheq}
\usepackage{makecell}
\usepackage{xcolor,colortbl}
\usepackage[ruled,vlined]{algorithm2e}
\usepackage{array,multirow}
\usepackage{comment}
\usepackage{pifont}
\usepackage{siunitx}
\usepackage{url}
\usepackage{enumitem}
\usepackage{arydshln}
\usepackage{xcolor}
\usepackage{soul}
\usepackage{graphicx}
\usepackage{graphicx}
\usepackage{caption}
\usepackage{url}
\usepackage{footnotehyper}
\makesavenoteenv{table}
\makesavenoteenv{tabular}

\newcommand{\nth}[1]{#1^{\text{th}}}
\newcommand{\bX}[1]{\mathbf{X}_{#1}}

\begin{document}

%

%

\runningtitle{IBM Research, Haifa}

\twocolumn[

\aistatstitle{Statistical multi-metric evaluation and visualization\\ of LLM system predictive performance}

\aistatsauthor{ Samuel Ackerman \And Eitan Farchi \And  Orna Raz \And Assaf Toledo }
\aistatsaddress{ IBM Research, Haifa } ]

\begin{abstract}
  The evaluation of generative or discriminative large language model (LLM)-based systems is often a complex multi-dimensional problem.  Typically, a set of system configuration alternatives are evaluated on one or more benchmark datasets, each with one or more evaluation metrics, which may differ between datasets.  We often want to evaluate---with a statistical measure of significance---whether systems perform differently either on a given dataset according to a single metric, on aggregate across metrics on a dataset, or across datasets.  Such evaluations can be done to support decision-making, such as deciding whether a particular system component change (e.g., choice of LLM or hyperparameter values) significantly improves performance over the current system configuration, or, more generally, whether a fixed set of system configurations (e.g., a leaderboard list) have significantly different performances according to metrics of interest. We present a framework implementation that automatically performs the correct statistical tests, properly aggregates the statistical results across metrics and datasets (a nontrivial task), and can visualize the results.  The framework is demonstrated on the multi-lingual code generation benchmark CrossCodeEval, for several state-of-the-art LLMs.
  


\end{abstract}

\section{Introduction
\label{sec:introduction}
}

There has been an explosion in recent years in the application of LLMs to solve a wide variety of complex tasks, such as code and natural language generation.  We use the term `LLM-based system' for any larger framework that uses LLMs as a component to perform a task, such as retrieval-augmented generation (RAG).  The rapid growth in the number and types of tasks that LLMs are applied to has been accompanied by a proliferation of benchmark datasets created to evaluate the performance of such systems.  Many leaderboards have been created which rank the performance of the newest LLMs on a given set of benchmarks, for the sake of transparency and competition.  For instance, HuggingFace's \href{https://huggingface.co/open-llm-leaderboard}{Open LLM Leaderboard} (see \cite{open_leaderboard_blog}) displays results for a set of LLMs on a set of benchmark datasets, as well as the average across datasets, which is used to assign an overall ranking to each model; such score aggregation is common but may be too simplistic.

LLM-based systems, similarly to ones involving standard machine learning algorithms, are often a pipeline consisting of multiple steps (e.g., pre- and post-processing of system inputs and outputs) and hyperparameters (e.g., choice of prompt template, retrieval method, and generation hyperparameters such as temperature and maximum token length).  Often, the choice of LLM itself can be an interchangeable component (i.e., hyperparameter) of the system as well.  This leads to a potentially large search space\footnote{Note, the current work does not try to optimally explore the search space of system hyperparameter values, as in \cite{ackerman2024using}, which conducts a factorial experiment.  Rather, we take the set of system configurations as given by the user.} for the best-performing system for different combinations of hyperparameter values (or the LLM itself).  Furthermore, one may want to use multiple evaluation metrics (criteria) and multiple benchmark datasets or tasks to decide on the best system hyperparameter values.

Leaderboards tend to display LLM scores and rankings \textit{without} an indication of the statistical significance of differences in scores.  More generally, one may want to know if, say, upgrading the LLM in an existing system to the newest version improves the average score by 0.02, whether this is a statistically significant improvement; such decisions can be very practical if the system is a component in a product.


In this work, we present a comprehensive framework and implementation\footnote{See \url{http:github.com/<hidden_for_anonymity>}.} for comparing multiple system configurations across multiple evaluation metrics and datasets, where the comparisons are done on a rigorous basis of statistical testing and significance.  This aspect of statistical significance is one that seems lacking in many existing attempts to evaluate models, such as the leaderboards.  Crucially, the testing is done `under-the-hood' with minimal user input, other than the underlying data.  The framework includes

\begin{itemize}[noitemsep]
    \item Support for statistical decisions on both paired and unpaired observation datasets.
    \item Proper statistical aggregation of metrics (criteria) for each system on a single dataset, and statistical comparisons thereof, and aggregation of system comparison results across evaluation datasets.
    \item Plotting utilities for exploratory visualization of system metric scores and statistical comparison results.
\end{itemize}

In short, the framework allows one to statistically evaluate and compare systems across metrics and datasets, and to plot the results.  In particular, this is done with as much ease to the user as possible, e.g., automatically determining the appropriate statistical test to use in each data setting.

We note that our statistical testing framework differs from approaches in the operations research field of multiple-criteria decision-making (MCDM; see e.g., \cite{stanujkic2013comparative}), in which a decision is made between multiple items (e.g., systems) across multiple criteria (e.g., evaluation metrics).  These techniques typically use a 2-dimensional items$\times$criteria matrix and mathematical optimization strategies, which may directly consider the individual influence of each metric in the decision, rather than formal statistical considerations.  On a single dataset of observations, our approach would be to directly aggregate metrics across observations and perform performs statistical comparisons on the systems on the basis of the aggregation, rather than directly doing an optimization across the multiple criteria.  In some cases, an MCDM approach may be better.  However, in many cases, using a metrics-aggregate as part of a wider evaluation across multiple benchmarks, which our framework does, with both flexibility and statistical-significance determination, is preferred, even if is not strictly a mathematically-optimized `decision' as in MCDM.

The structure of the paper is as follows.  Section~\ref{sec:terminology_and_notation} introduces the basic terminology and notation used, with a motivating example.  Section~\ref{sec:exploratory_visualization} illustrates the exploratory plotting utilities (prior to any statistical testing) that the framework includes.  Section~\ref{sec:statistical_testing} introduces the statistical comparison of systems.  Section~\ref{sec:visualizations} illustrates the visualization of statistical testing results on a metric.  Section~\ref{sec:aggregation} explains the aggregation of metrics on a dataset, and of statistical results across datasets.  Section~\ref{sec:cceval_analysis} illustrates a complete evaluation of systems, using aggregation across datasets.  Section~\ref{sec:conclusion} concludes.  The supplementary materials discuss auxiliary utilities in our implementation that aid system testing.

Throughout, our motivating example is CrossCodeEval \citep{ding2024crosscodeeval}, a multi-lingual code completion benchmark.  Our analysis will use results on this dataset from \cite{mishra2024granite}.  While our work is motivated by the application of LLM evaluation, the framework operates only on the post-generation evaluation results.  Thus, it is equally applicable to evaluating non-LLM systems as well.
\section{Terminology and notation
\label{sec:terminology_and_notation}
}

\subsection{Basic units of analysis
\label{ssec:basic_units}
}

Here, we will define notation for the data objects to be analyzed.

A \textbf{dataset} $D$ is a fixed set of $n$ examples of inputs $x_i, i=1,\dots,n$; often, each $x_i$ has a corresponding reference (ground truth) value $y_i$.  For instance, in CrossCodeEval, there is a separate code completion dataset for each programming language; there, each $y_i$ is the true missing code and $x_i$ is the surrounding code.

A \textbf{system} is a function $f$ that returns an output $\hat{y}_i=f(x_i)$ on dataset inputs.  For instance, in a generative setting $\hat{y}_i$ can be generated code completion based on inputs $x_i$, or translation of $x_i$ into another language; if a reference $y_i$ exists, a good output $\hat{y}_i$ should be close to $y_i$.  $\hat{y}_i$ can also be discriminative or predictive output, such as in binary question-answering or part-of-speech tagging.  In our case, a system is defined by the entire specification of components such as choice of LLM, retrieval method, prompt template, generation hyperparameters, etc.; any change in components constitutes a different system.

A \textbf{metric} $m$ is a function $m(\hat{y}_i,\dots)$ that scores outputs $\hat{y}_i$.  If $D$ has reference values, typically the metric takes the form $m(\hat{y}_i, y_i)$ and scores the outputs vs the references (e.g., edit similarity, precision), but it can also be a reference-less evaluation $m(\hat{y}_i)$ like summary coherence or length, or may even incorporate $x_i$.  Metrics (or transformations of them) should positively or negatively \textit{monotonically} reflect measured output quality.  For instance, a similarity metric $m(\hat{y}_i,y_i)$ will be higher if the quality is better relative to the reference, whereas the opposite is true for distance metrics.

A \textbf{score sample} $V=(v_1,\dots,v_n)$ is a set of $n$ metric scores $v_i=m(\hat{y}_i,\dots)$. Each sample $V$ corresponds to a specific dataset, metric, and system.

A \textbf{score sample list} $L=(V_1,\dots,V_B)$ is a set of score samples that can be validly compared: each $V_b\in L$ is a score sample corresponding to a \textit{different system} $b$ (out of $B$ total), but using the \textit{same} dataset and metric.

A \textbf{score sample list collection} $C=(L_1,L_2,\dots)$ is a set of score sample lists $L_i$.  The lists are mutually compatible if each measures the same set of $B$ systems, but on a different metric-dataset combination.

\subsection{Observational pairing
\label{ssec:pairing}
}
In general, two samples of values $(v_1,\dots,v_n)$ and $(z_1,\dots,z_m)$ are considered \textbf{paired} if $m=n$ and each $(v_i,z_i)$ represent measurements on the same observational unit $i$; otherwise they are \textbf{unpaired}.  The pairing determines the statistical comparisons (Section~\ref{sec:statistical_testing}) that can be done.  In our framework, one creates a testing object for a defined set of systems, that expects data that is either paired or unpaired. In paired analysis, all score samples $V_b$ in a score sample list $L$ must be from the same fixed dataset $D$.  If samples are allowed to be unpaired, $V_b\in L$ can be of different sizes.

\begin{figure}
\centering
\includegraphics[width=0.95\columnwidth]{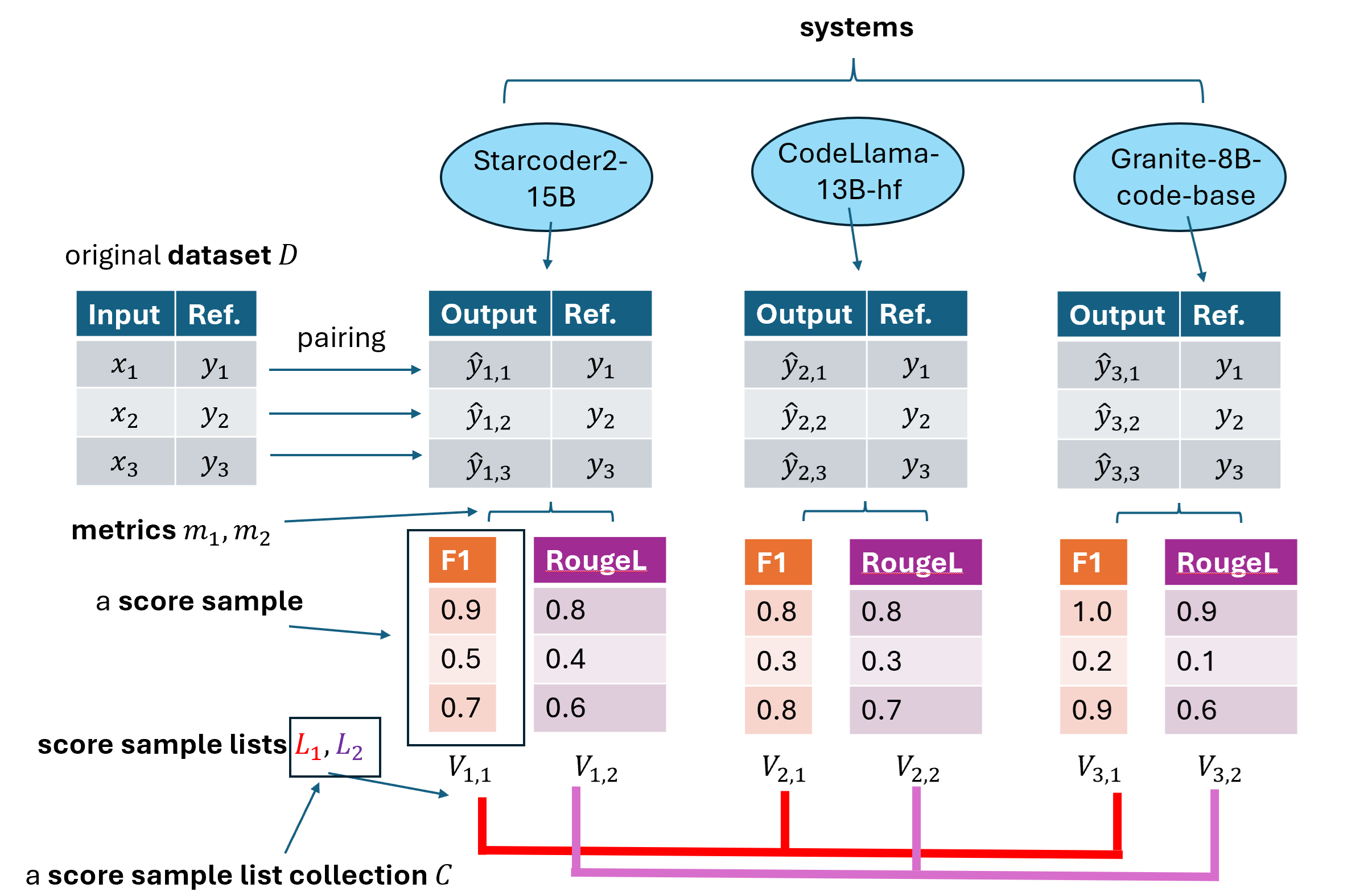}
\caption{Illustration of terms (Section~\ref{ssec:basic_units}) for paired data.}
\vspace{-0.1in}
\label{fig:notation_diagram}
\end{figure}

Figure~\ref{fig:notation_diagram} illustrates the terminology of Section~\ref{ssec:basic_units} for the paired data setting. 
\begin{itemize}[noitemsep]
    \item A \textbf{dataset} $D$ of $n=3$ inputs $x$ and (optional) references $y$.
    \item The blue bubbles are $B=3$ \textbf{systems}, labeled by the name of the LLM; for instance, the system could all use the same components other than the LLM.  The ``Output" columns are the outputs $\hat{y}_i$ for each input $x_i$.
    \item The two \textbf{metrics} used are F-1 and Rouge-L scores.
    \item Each of the F-1 and Rouge-L tables is a \textbf{score sample} of size $n$.  The leftmost is $V_{1,1}=(0.9, 0.5, 0.7)$, corresponding to the 1st system using the 1st metric.
    \item Collectively, each of the F-1 and Rouge-L score samples forms a \textbf{score sample list} of size $B=3$, as indicated by the brackets. These can be denoted $L_1=(V_{1,1},V_{2,1},V_{B,1})$ for F-1, and $L_2=(V_{1,2},V_{2,2},V_{B,2})$ for Rouge-L.  $L_1$ and $L_2$ are mutually compatible for testing since their metrics are measured on the same $B=3$ systems.  In this case they are also on the same dataset (are paired), but that is not required. 
    \item $C=(L_1,L_2)$ forms a \textbf{score sample list collection}.  
\end{itemize}

\section{Exploratory visualization
\label{sec:exploratory_visualization}
}

All the analysis conducted in this work is shown in the jupyter notebook \texttt{cceval\_analysis}.  We first show an example of visual exploratory analysis that does not involve formal statistical testing.

For a score sample list $L=(V_1,\dots,V_B)$ of a single metric, for each system $b$ we can plot the distribution of either
\begin{enumerate}[noitemsep]
    \item the observed metric values $V_b$ themselves,
    \item the population mean of the distribution of $V_b$, obtained by bootstrapping, or
    \item the system ranks (the best system receives a rank value of 1, the second-best 2, etc.).
\end{enumerate}

using boxplots.  The utility also can display confidence intervals---generally less informative about the distribution shape---for the $100(1-\alpha)\%$ confidence level (default 95\%) specified by the tester object (see Section~\ref{sec:statistical_testing}).  Simultaneous confidence intervals for ranks are calculated using the Tukey-interval-based method in \cite{al2022simultaneous}; the rank distributions are estimated by bootstrapping (either paired or unpaired) of the score samples, and calculating the ranks of the resampling means.  

\begin{figure}[h!]
\centering
\includegraphics[width=0.95\columnwidth]{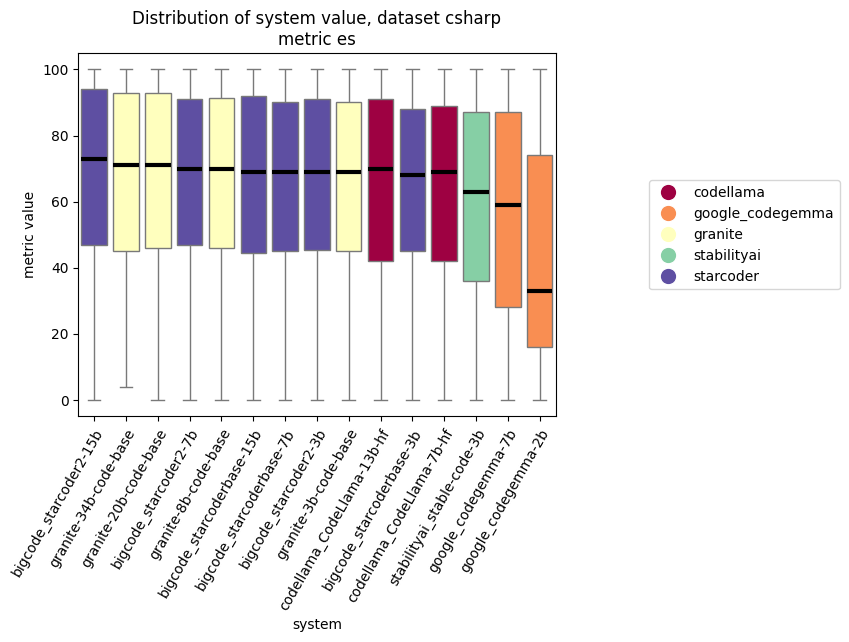}
\includegraphics[width=0.95\columnwidth]{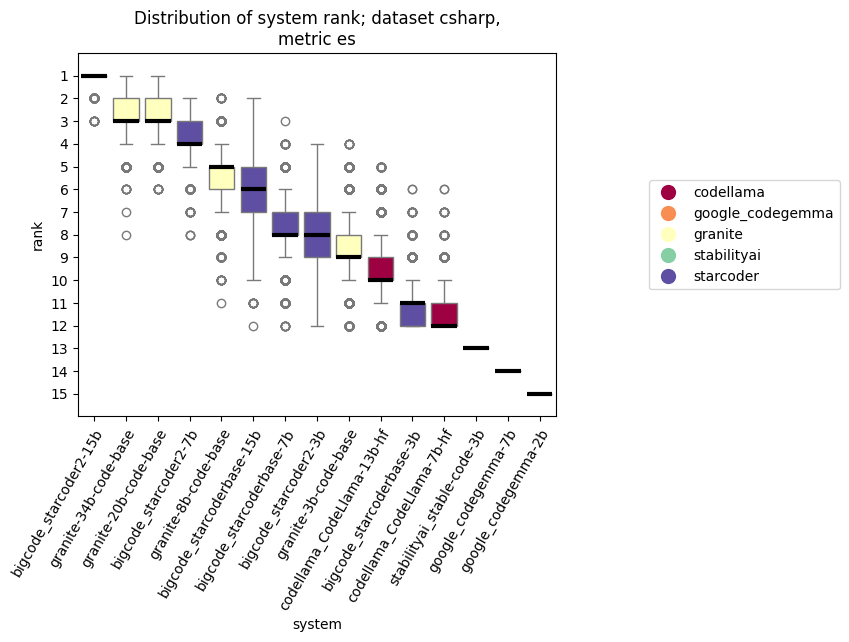}
\caption{Boxplots of system value and rank distribution for the ES metric on CrossCodeEval's C\# dataset. Systems are ordered from left to right in decreasing mean quality (for value distribution) or decreasing median rank.}
\vspace{-0.1in}
\label{fig:boxplots}
\end{figure}

Figure~\ref{fig:boxplots} shows examples of boxplots of options 1 and 3 above for the edit similarity (ES) metric on the C\# language dataset, where systems are ordered from left to right by decreasing quality.  In this case, the rank distributions (bottom) can be helpful in indicating that the five best systems have high probability of being differentiated into ranks 1, 2, 3, etc., since the rank distributions are separable but the metric value boxplots themselves look very similar.  Thus, the ranks can assure a user that the system with the best score average will in fact with high probability be the best.
We note also that the utility allows systems to be colored automatically by a user-specified grouping. 

\section{Statistical testing
\label{sec:statistical_testing}
}

\subsection{Choice of tests
\label{ssec:choice_of_tests}}

Let $V_i,V_j$ be score samples corresponding to systems $i$ and $j$; they can be compared if they are from the same sample list (see Section~\ref{ssec:basic_units}), meaning they represent values of the same metric on the same dataset $D$, with the same pairing type.  We can evaluate the systems' relative performance by comparing the distributions of samples $V_i,V_j$, depending on whether a higher metric value is better.

It is common that data analysts may not know the proper statistical test to perform in a given scenario.  Our utility automates this, with the test performed depending on the metric modality (binary or numeric-valued) and the pairing, as shown in Table~\ref{tab:statistical_tests}.

\begin{table}[ht]
\begin{center}
\begin{tabular}{l l | l l}
\hline
paired & modality & hypothesis test & effect size\\\hline
 no & numeric & Welch's T-test\footnote{An independent-samples test without the assumption of equal population variances; see \href{https://docs.scipy.org/doc/scipy/reference/generated/scipy.stats.ttest_ind.html}{\textrm{scipy.stats.ttest\_ind}} \citep{2020SciPy-NMeth}.} & Cohen's\footnote{For two numeric-valued univariate samples $\bX{1}$ and $\bX{2}$ of sizes $n_1,n_2$, defined as $d=\frac{\overline{\bX{1}} - \overline{\bX{2}}}{s_P}$, where the pooled sample standard deviation is $s_P=\sqrt{\frac{(n_1-1)\textrm{Var}(\bX{1}) + (n_2-1)\textrm{Var}(\bX{2})}{n_1+n_2-2}}$; see \citealt{C1988}[pp. 66--67].} $d$\\
 no & binary & Z proportions\footnote{\href{https://www.statsmodels.org/dev/generated/statsmodels.stats.proportion.test_proportions_2indep.html}{Independent proportions Z-test} \cite{seabold2010statsmodels} tests whether the two binary-valued samples have equal (proportion-valued) population means.} & Cohen's\footnote{For two proportions $p_1,p_2\in[0,1]$, $h$ is defined as $2\arcsin{(\sqrt{p_1})}-2\arcsin{(\sqrt{p_2})}$ (\citealt{C1988}[pp. 179--182]).} $h$\\ 
 \hline
 yes & numeric & paired T-test\footnote{See \href{https://docs.scipy.org/doc/scipy/reference/generated/scipy.stats.ttest_rel.html} {\textrm{scipy.stats.ttest\_rel}}, \cite{2020SciPy-NMeth}.} & paired Cohen's\footnote{For paired numeric-valued univariate samples $\bX{1}$ and $\bX{2}$ of size $n$, let vector $\boldsymbol{\Delta}=\bX{1} - \bX{2}$.  Defined as $d=\frac{\overline{\boldsymbol{\Delta}}}{\textrm{StdDev}(\boldsymbol{\Delta})}$; see \citep{C1988}[pp. 48].} $d$\\
 yes & binary & McNemar's test\footnote{An exact test for binary data calculated on a $2\times 2$ contingency table.  See \href{https://www.statsmodels.org/dev/generated/statsmodels.stats.contingency_tables.mcnemar.html}{McNemar's test}, \cite{seabold2010statsmodels}.} & paired Cohen's\footnote{Used instead of binary-specific effect sizes like Cohen's $g$ (see \citealt[pp. 147--150]{C1988}; \citealt[`Tests for Paired Nominal Data
']{mangiafico2016summary}) which can be calculated from a $2\times 2$ contingency table, because $d$ is directed, unlike Cohen's $g$, and can be seamlessly aggregated with the other effect sizes.
} $d$\\
 \hline
\end{tabular}
\end{center}
\caption{\label{tab:statistical_tests}Statistical hypothesis test and effect size metric for each combination of observation pairing and data modality.  Permutation tests for each can be performed instead, but then effect sizes are not calculated.}
\end{table}

Importantly, for each sample comparison $(i,j), i\ne j$, our utility returns not only the p-value $p_{i,j}$ of the appropriate hypothesis test, but also an effect size value $e_{i,j}$.  Effect size metrics are often recommended to accompany the more familiar p-value because effect sizes are considered to more accurately reflect the `practical' magnitude of a given effect (e.g., difference in means) and are resistant to becoming significant simply due to large sample sizes (`p-hacking'), which is a problem with p-values; see \cite{sullivan2012using} for a discussion.

A comparison p-value $p_{i,j}$ is determined to be significant if $p_{i,j}<\alpha$, where $\alpha$ (default 0.05) is specified by the user for the tester object.  Effect sizes---many of which are discussed in \cite{C1988}---are determined to be significant if they exceed the respective numeric threshold constituting either a "medium" or "large" difference, as chosen by the user; see \cite{sawilowsky2009new}.

\subsection{Order of comparisons
\label{ssec:order_of_comparisons}
}

For a score sample list $L=(V_1,\dots,V_B)$ of score samples for systems $1,\dots,B$, our utility conducts ordered pairwise comparisons $(i,j)$ in the set $\Omega$ in one of three ways:
$$
\Omega=\begin{cases}
    \{(i,j)\colon\:i=1,\dots,B-1; j=i+1,\dots,B\},\textrm{ or}\\
    \{(1,j)\colon j=2,\dots,B\},\textrm{ or}\\
    \{(i,i+1)\colon i=1,\dots,B-1\}	 \end{cases}
$$
These options are, respectively, to compare \textit{all pairs} of systems (the default), the \textit{first vs the rest}, or \textit{successive pairs} of systems.  Tests are done either with a two-sided (default), or left or right-sided alternative.  The non-default comparisons $\Omega$ and alternatives should only be used if the user has pre-ordered systems $1,\dots,B$ in order of expected quality, without first examining the metric values, which would make the p-value false-positive probability guarantee invalid.  For instance, one could order the systems by decreasing number of parameters (a reasonable proxy for system power) or by decreasing perceived relevance to the task at hand of the systems' pre-training datasets.

The testing procedure returns p-values $(p_{i,j})$ and effect sizes $(e_{i,j})$ for $(i,j)\in \Omega$ as chosen.  The p-values used for all analyses are after adjusting the raw p-values $(p_{i,j})$ for multiple comparisons\footnote{\href{https://www.statsmodels.org/dev/generated/statsmodels.stats.multitest.multipletests.html}{\texttt{statmodels}'s \texttt{multipletests}}, \cite{seabold2010statsmodels}.} using the (default) Holm-\v{S}\'{i}d\'{a}k step-down procedure.  This procedure is known to control the familywise error rate (FWER) at the chosen $\alpha$ for any dependence structure of tests; hence, it is appropriate for all options\footnote{Existing methods adjust p-values for the cases of `all pairs' (\href{https://docs.scipy.org/doc/scipy/reference/generated/scipy.stats.tukey_hsd.html}{Tukey HSD test}) and `first vs the rest' (\href{https://docs.scipy.org/doc/scipy/reference/generated/scipy.stats.dunnett.html}{Dunnett's test}) comparisons, but these require the limiting assumptions of normally-distributed independent (and not paired) samples with equal population variances.} of $\Omega$.  We note that effect sizes are not similarly adjusted for multiplicity, and that often differences that are significant by the p-value criterion do not have very significant effect sizes.

Many users may use the simplest system evaluation strategy of selecting the system with the best performance\footnote{Highest value of (aggregate) metric mean $\dot{V}_b$, see Section~\ref{ssec:post_op_plotting_procedure}.} out of the $B$ considered.  This system's performance would then be tested only against the second-best, which does not require multiplicity adjustment; it would be selected regardless of the statistical significance of the test, which merely informs the decision, because even small improvements that are not statistically significant may be important. Such a strategy would not require many of abilities our framework provides. 

However, our framework is designed to give users maximal insights to inform their decision, even if they may ultimately use only this simple strategy.  Visualizing the groupings (cliques, see Section~\ref{ssec:connected_graph}) of full pairwise comparisons of the $B$ systems can reveal useful insights, for instance that the third-best system is much less computationally heavy than the two best, but is not statistically significant from them (they form a clique).  This insight may lead the user to select \textit{that} system rather than the best because there is another consideration.  Furthermore, our framework is general to any predictive system, not just LLMs, and users in other domains may have system selection strategies (e.g., placing a larger emphasis on statistical significance or other system aspects) that differ from the simple one here. 
A one-size-fits-all-domains strategy is not always appropriate. \section{Statistical visualizations
\label{sec:visualizations}
}

Performing pairwise statistical tests on a (single-metric) score sample list, as in Section~\ref{sec:statistical_testing}, returns a formatted test report object.  This object can be passed to the visualization utilities discussed here; also, multiple compatible report objects can be aggregated into a new aggregate report object (Section~\ref{ssec:test_result_aggregation}), and similarly used.

\subsection{Connected graph plot
\label{ssec:connected_graph}
}

This visualization is an original contribution.  It displays a score sample list test result as a connected graph $G=(V,E)$, where vertices $V=(1,\dots,B)$ correspond to the $B$ systems, and edges $E$ correspond to the pairwise significances (p-value or effect sizes).  Here and throughout, the p-values are adjusted for multiplicity (Section~\ref{ssec:order_of_comparisons}).  A vertex's vertical position corresponds to the sample average metric value for that system, with the highest ones being best; the horizontal position is meaningless.  The best system is labeled in red boldface, and therefore is the recommended system choice, though not necessarily with a statistically significant difference from the next-best.  $G$ is directed if the test alternative hypothesis is left/right-tailed, otherwise undirected.  Edges only potentially exist between vertex pairs in $\Omega$; an edge is drawn only if the comparison is \textit{in-significant} according to the p-value or effect size criterion, with the edge thickness increasing with in-significance.   

One can see quickly which systems---the highest vertices---are best on average.  A system that is significantly different than others will have few, if any, edges.  If many systems are statistically similar, the graph will have many thick edges.  Thus, a graph in which the highest vertices are not connected to others is a clear indicator that these systems are significantly the best.

\begin{figure}
\centering
\includegraphics[width=0.91\columnwidth]{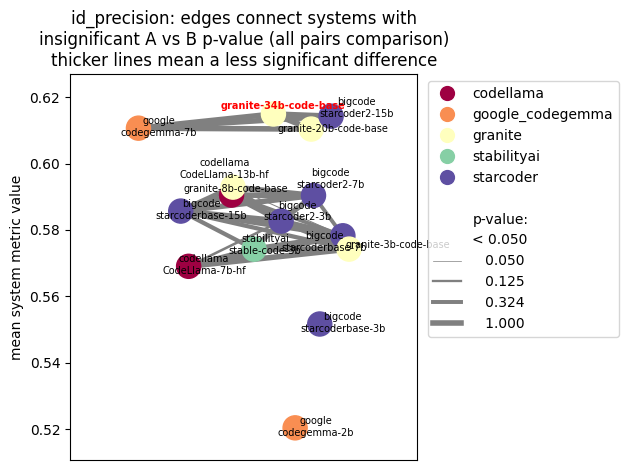}
\includegraphics[width=\columnwidth]{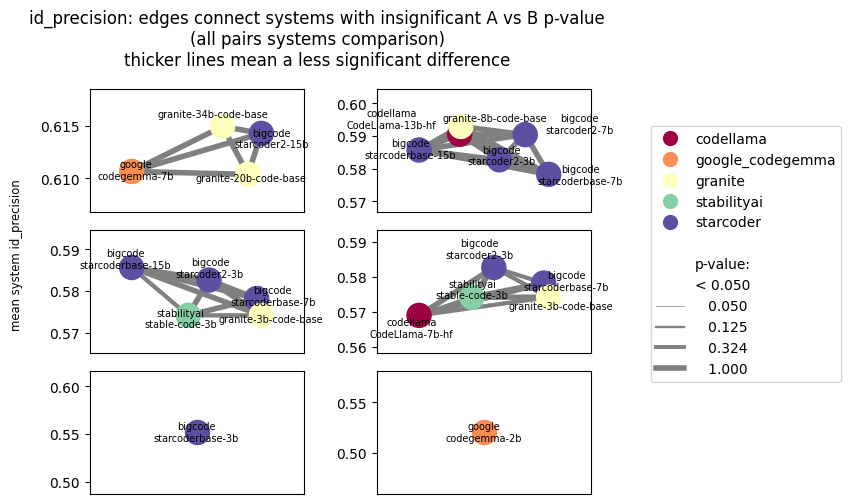}
\caption{Top: Connected graph using p-values for ID-precision metric on CrossCodeEval's Python dataset.
\\Bottom: The graph split into system cliques of at least 2.
}
\vspace{-0.1in}
\label{fig:connected_graph}
\end{figure}

This plot scales better than the heatmap (below) when the number of systems grows.  However, the visualization can become busy if there are many edges.  Thus, if `all pairs' $\Omega$ comparisons are done, there is an option to additionally find all cliques of $G$ of a minimum size.  Systems that appear together in a clique (which can overlap) can be considered a group that all have statistically similar performance.  Then, all cliques and isolated vertices are plotted in descending mean metric value (clique vertex height).  Figure~\ref{fig:connected_graph} shows an example using the p-value significance.  See Figure~\ref{fig:cceval_full_connected_graph} for an example of cliques.

\subsection{Heatmap table
\label{ssec:heatmap_table}
}

This plot can display test results for a score sample list collection of multiple metrics.  Each row corresponds to a comparison system pair $(i,j)\in \Omega$, and each column to a metric (a list).  Cells that are white are not significant comparisons for that metric; otherwise, the cell displays the significance value (p-value or effect size, as chosen).  Rows are ordered so that pairs appear in descending aggregate significance (see Section~\ref{ssec:test_result_aggregation}) across the metrics in the collection.  Cell backgrounds are colored so that darker colors are more significant, with red/blue colors indicating the direction.  Thus, if all metrics have the same inherent directionality, a row with both red and blue cells indicates that the metrics conflict in their conclusions about the best system in the pair.

\begin{figure}
\centering
\includegraphics[width=0.91\columnwidth]{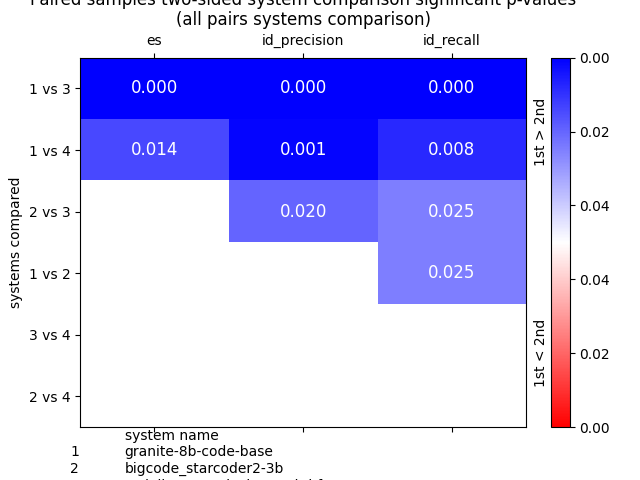}
\caption{Heatmap of p-values for all metrics for CrossCodeEval's Python dataset, for four systems.}
\vspace{-0.1in}
\label{fig:heatmap_graph}
\end{figure}

The heatmap table can be used to easily show the most significantly different pairs of systems (at the highest rows) and which metrics yield more or less significant differences (by the cell coloring).  The table is useful to display the raw test result values, but it can become unwieldy if the number of pairwise comparisons (table rows), which grows quadratically with the number of systems if all pairs are tested, is too large.  See Figure~\ref{fig:heatmap_graph} for an example on a subset of four systems.

\subsection{PairWise P-values Plot (PWPP)
\label{ssec:paired_pvalues}
}

This visualization is an adaptation of the \texttt{pwpp} plot in the \texttt{R} package \texttt{emmeans} (\cite{emmeans}; see \cite{emmeans_vignette}).  The systems (vertical axis) in each compared pair in $\Omega$ (Section~\ref{ssec:order_of_comparisons}) are connected by a vertical line whose horizontal location corresponds to the adjusted p-value $p_{i,j}$ (not the Tukey p-value in \texttt{emmeans}).  The horizontal axis scale is skewed so that more space is given to the area of significance $<\alpha$, to emphasize these p-values.  The concentration of line segments on the left side of the plot indicates which systems were frequently different from others.
\section{Aggregation
\label{sec:aggregation}
}
Having discussed visualizations of metric score sample distributions and statistical tests on them, we explain how these can be aggregated for further analysis.

Consider a score sample list collection (Section~\ref{ssec:basic_units}) $C=(L_1,\dots,L_K)$, where each score sample list is $L_j=(V_{1,j},\dots,V_{B,j})$.
For now, assume all systems are scored on a single dataset $D$, so that list $L_j$ corresponds to the $\nth{j}$ metric, and $V_{b,j}$ is a score sample for the $\nth{b}$ system scored by the $\nth{j}$ metric.
Let $\bar{V}_{b,j}$ and $S_{b,j}$ be the observed sample mean and standard deviation for $V_{b,j}$.  Let $\bar{V}_{\cdot,j}$ and $S_{\cdot,j}$ be the corresponding statistics of the pooled system score samples in $L_j$ for the $\nth{j}$ metric.  Let $n_{b.j}=|V_{b,j}|$ be the score sample sizes, which are all equal in the fully paired setting. Let $I_j=1$ if a higher value of metric $j$ indicates better performance, otherwise $-1$ if it is \textit{worse}.

\subsection{Score sample aggregation
\label{ssec:sample_aggregation}
}

Assume first that the observations are all paired on the dataset $D$; i.e., each system is scored on a fixed dataset $D$, rather than on (possibly unequally-sized) different random samples from a larger `population' dataset $D$.  

In the paired case, for each system $b$, score samples $V_{b,1},\dots,V_{b,K}$ can be aggregated directly across metrics $1,\dots,K$.  First, observation values in $V_{b,j}$ are standardized by the pooled metric mean $\bar{V}_{\cdot,j}$ and standard deviation $S_{\cdot,j}$ so that the pooled scores for each metric $j$ have mean 0 and standard deviation 1; this rescaling is to ensure that the scores for each metric, which may have inherently different value ranges, can be compared on a common scale.  The now-standardized values $\tilde{V}_{b,j}$ are multiplied by $I_j$ so that higher values always are better.  Then, for each system $b$, $\tilde{V}_{b,j}$ 
are averaged element-wise across the metrics---possibly giving unequal weights to the metrics according to user sense of metric importance---yielding an aggregate score sample list $L_{\textrm{agg}}=(V_{1,\textrm{agg}},\dots,V_{B,\textrm{agg}})$.  This new metric is an aggregate of the $K$ original ones, and can be used in significance tests (Section~\ref{sec:statistical_testing}) or visualization (e.g., Section~\ref{sec:exploratory_visualization} and \ref{sec:visualizations}) as any score samples list.  Because the metric-wise averaging adjusted the direction of each metric value by $I_j$, for the aggregate average, higher is also always better.

\subsection{Test result aggregation
\label{ssec:test_result_aggregation}
}

Assume now that in the collection $C$, the score sample lists $L_j$ are not observation-paired.  For instance, the $L_j$ are from different random draws of the same dataset, or from different datasets; the metrics could be the same or different for each dataset.  In CrossCodeEval (see Section~\ref{ssec:cceval_analysis_procedure}), $L_j$ could be the individual or aggregated metrics, each on a different programming language dataset $j$.

Because the direct aggregation preserves the most information of the data, it is recommended to directly aggregate as far as the pairing allows, for instance from individual metrics to an aggregate one on the same dataset.  But if the score samples $L_j$ are no longer paired across different $j$, the values cannot be directly aggregated to give a single sample score list that can be tested or visualized.  Rather, we can aggregate the result of statistical tests, as follows:
\begin{enumerate}
    \item For each score sample list $L_j$, conduct pairwise tests 
    of system pairs $(a,b)\in \Omega$ (Section~\ref{ssec:order_of_comparisons})  score samples $(V_{a,j},V_{b,j})$.
    \item For each such test $j$, return the appropriate (adjusted) p-value $p_{a,b,j}$ and effect size value $e_{a,b,j}$ (Table~\ref{tab:statistical_tests}).
    \item For each system pair $(a,b)$, calculate p-values $p_{a,b,\cdot}$ and effect sizes $e_{a,b,\cdot}$, aggregated across score sample lists $L_j$ (i.e., metrics $j$), as described below.
\end{enumerate}

For each compared system pair $(a,b)$, the raw per-metric p-values $p_{a,b,j},\:j=1,\dots,K$ are aggregated into a single p-value $p_{a,b}$ using Wilson's harmonic mean\footnote{Unlike p-value combination methods implemented in \texttt{scipy.stats.combine\_pvalues} (see \cite{heard2018choosing} for a discussion), it better reflects the average of the p-values; Stouffer's method, for instance, returns a final p-value of 1 if any of the individual p-values are 1.} p-value (HMP) method \cite{wilson2019harmonic}.  The adjusted p-value, denoted $\mathring{p}_{a,b}$ by Wilson, accounts for the total $L=|\Omega|\times K$ pairwise tests performed (see \cite{harmonicmeanp_tutorial}), and controls the FWER, optionally using metric weights $w_j$.  Because the metric aggregate p-values are not raw, aggregated results cannot be further aggregated.


Effect sizes are aggregated following \cite{turner2006calculating} by using the standard deviation of the data underlying the effect size as inverse weights in averaging.  The inverse weights are used so that samples whose effect size is less variable receive higher weight.  For each effect size $e_{a,b,j}$, the standard deviation is the denominator of Cohen's $d$ (paired or unpaired as appropriate) calculated using the standardized metric values $\tilde{V}_{a,j},\tilde{V}_{b,j}$ (see Section~\ref{ssec:sample_aggregation}).  Because the interpretation thresholds for the various Cohen effect sizes metrics are the same, the inversely-weighted aggregate effect size can be evaluated for statistical significance according to the same thresholds.

\subsection{Unpaired score sample mean aggregation
\label{ssec:unpaired_mean_aggregation}
}

In the unpaired setting, we can calculate standardized system metric aggregate means $\dot{V}_b$ across the collection $C$ of unpaired datasets; recall that in unpaired data, observation scores cannot be aggregated directly as in Section~\ref{ssec:sample_aggregation}, and further we assume that we have only summary statistics on $C$ and not the full samples.  For each system $b=1,\dots,B$, we calculate a standardized mean $\dot{V}_b$ as follows:

\begin{enumerate}
    \item Consider the observed score sample mean $\bar{V}_{b,j}$ as a random variable of the mean observed on a sample of size $n_{b,j}$ with standard deviation $S_{b,j}$.
    \item For score sample list $L_j$, calculate the pooled standard deviation $\hat{S}_{\cdot,j}=\sqrt{\frac{\sum_{b=1}^B(n_{b,j}-1)S^2_{b,j}}{(\sum_{g=1}^{B} n_{g,j})-B}}$, which approximates $S_{\cdot,j}$.   Standardize the observed $(\bar{V}_{1,j},\dots,\bar{V}_{B,j})$ using the standard deviation $\hat{S}_{\cdot,j}\sqrt{\sum_{b=1}^B \frac{1}{n_{b,j}}}$, which is the denominator in an independent-samples T-test (Table~\ref{tab:statistical_tests}), yielding the values $\tilde{\bar{V}}_{b,j}$. 
    \item Let $\dot{V}_{b}=\sum_{j=1}^Jw_jI_j\tilde{\bar{V}}_{b,j}$, where $w_j$ are weights summing to 1, the same used for p-values in Section~\ref{ssec:test_result_aggregation}.
\end{enumerate}

\section{Example of analysis on CrossCodeEval
\label{sec:cceval_analysis}
}

As noted in Section~\ref{sec:introduction}, we will use results from \cite{mishra2024granite} of evaluations of 15 LLM systems on four programming language datasets (C\#, Java, Python, and Typescript) of the CrossCodeEval benchmark \citep{ding2024crosscodeeval}.  Here we will use the three evaluation metrics edit similarity (ES), ID-precision, and ID-recall; ID-F1 is omitted since it is already a direct combination of the precision and recall.

\subsection{Analysis plotting procedure
\label{ssec:post_op_plotting_procedure}
}

Our framework contains a plotting method \texttt{post\_op\_ordered\_ranking} that performs the following procedure, given a score sample list collection $C$ (see Section~\ref{ssec:test_result_aggregation}), where $|C|>1$:

\begin{enumerate}
    \item \textbf{Perform aggregation} to a score sample list $L_{\textrm{agg}}$:
    \begin{enumerate}
        \item If the score sample lists $L_1,\dots,L_{|C|}$ are all mutually \textit{paired}, the aggregation follows Section~\ref{ssec:sample_aggregation}.  Calculate system means as $\dot{V}_b=\bar{V}_{j,\textrm{agg}}$.
        \item Otherwise, if \textit{unpaired}, perform pairwise tests for each score sample list $L_j$, then aggregate, following Section~\ref{ssec:test_result_aggregation}.  Calculate system means as $\dot{V}_b$ following Section~\ref{ssec:unpaired_mean_aggregation}.
    \end{enumerate}
    \item Order systems $1,\dots,B$ in descending order of $\dot{V}_1,\dots,\dot{V}_B$.
    \item Perform pairwise testing according to any of the comparisons $\Omega$ (Section~\ref{ssec:order_of_comparisons}).  If `all pairs', perform two-sided tests; otherwise, perform right-tailed tests since the systems means are in descending order.
    \item Display a connected graph plot (Section~\ref{ssec:connected_graph}), optionally showing cliques.
\end{enumerate}

\subsection{Analysis procedure on CrossCodeEval 
\label{ssec:cceval_analysis_procedure}
}

The programming language datasets are denoted $D_1,D_2,D_3,D_4$.  The datasets contain different sets of inputs $(x_i)$, and are of differing sizes, hence an analysis across them is unpaired.  However, for each $D_j$, because each system returns results on all observations within $D_j$, these analyses are paired.  Thus, we have the following procedure:

\begin{enumerate}
    \item \label{enum:agg_step} For each dataset $D_j,\:j=1,\dots,4$, create aggregate score sample list $L_{j,\textrm{agg}}$ (Section~\ref{ssec:sample_aggregation}) where each of its constituent $B=15$ score samples $V_{b,\textrm{agg}}$, $b=1,\dots,B$, is a direct (possibly weighted) aggregate of the three metric score samples for that system.
    \item Form the score sample list collection $C=(L_{1,\textrm{agg}}, L_{2,\textrm{agg}}, L_{3,\textrm{agg}}, L_{4,\textrm{agg}})$.  Since now the $L_{j,\textrm{agg}}$ across language datasets $j$ are \textit{unpaired}, follow the procedure\footnote{The procedure could have also received the full set of 12 un-aggregated score samples as a collection $C$, but it is preferable to first exploit the pairing within datasets to directly aggregate score sample values rather than only aggregating the full 12 score sample significance test results as in Section~\ref{ssec:test_result_aggregation}.} in Section~\ref{ssec:post_op_plotting_procedure}.  The aggregate metric created in the procedure is a new aggregate across languages of the metric aggregates within each language.
\end{enumerate}

\subsection{Analysis results on CrossCodeEval 
\label{ssec:cceval_analysis_results}
}

\begin{figure}[h!]
\centering
\includegraphics[width=0.92\columnwidth]{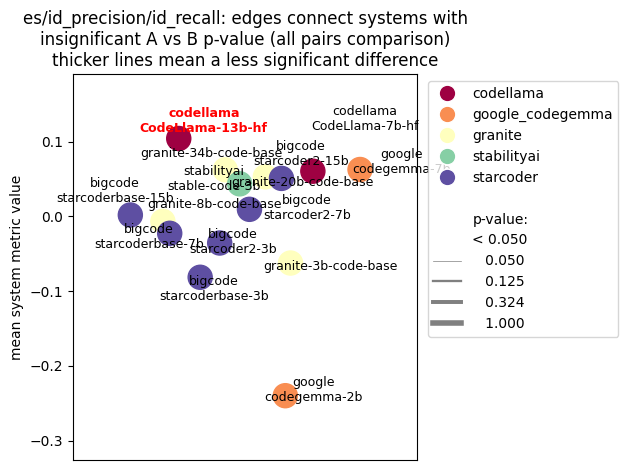}
\caption{Connected graph using p-values on CrossCodeEval, aggregated across program language datasets.
}
\vspace{-0.1in}
\label{fig:cceval_full_connected_graph}
\end{figure}

The procedure on CrossCodeEval, using full pairwise comparisons for the datasets and metrics under consideration, is shown in Figure~\ref{fig:cceval_full_connected_graph}.  By the p-value criterion (aggregated across languages using the adjusted harmonic mean, see Section~\ref{ssec:test_result_aggregation}), all system pairs are statistically significantly different from each other.

Table~\ref{tab:top_systems} shows the top six models according to our aggregate ranking method (see $\dot{V}_b$ in Section~\ref{ssec:post_op_plotting_procedure}).  They are CodeLlama-13B-hf, Google CodeGemma-7B, CodeLlama-7B-hf, Stable Code 3B, and Granite 34B- and 20B-code-base systems.  As noted in Section~\ref{sec:introduction}, a common, but naive, practice for aggregation is, for each system, to compute the overall average across datasets of the average metric average within each dataset.  The three best systems according to this averaging method, as shown in Table~\ref{tab:top_systems}, are the same as under our method; it ranks the Granite 34B and 20B as $\nth{4}$ and $\nth{5}$, with Stable Code as $\nth{6}$.  Thus, the same six systems are ranked best under both methods, with slight changes in ordering.  However, simple averaging does not indicate whether the differences in the averages are statistically significant, which our method does, among other advantages.


\begin{table}[ht]
\begin{center}
\begin{tabular}{l|rr}
\hline
system & averaging & our method ($\dot{V}_b$) \\
 \hline
CodeLlama-13B-hf & 0.5926 & 0.1345 \\
Google CodeGemma-7B & 0.5783 & 0.1086 \\
CodeLlama-7B-hf & 0.5768 & 0.0868 \\
StabilityAI Stable Code 3B & 0.5705 & 0.0744 \\
Granite-34B-code-base & 0.5749 & 0.0521 \\
Granite-20B-code-base & 0.5714 & 0.0418 \\
\hline
\end{tabular}
\end{center}
\caption{\label{tab:top_systems}Top 6 systems under simple aggregate-averaging, and our method.  Metric ES has been pre-divided by 100 to match the range of the other metrics.}
\end{table}

\subsection{Comparison with MCDMs
\label{ssec:mcdm_comparison}
}

To compare the results of our program-language metric aggregation with MCDM algorithms, we construct matrix $M_{B\times 3}$
\[
M=\begin{bmatrix}\bar{V}_{1,1} & \bar{V}_{1,2} & \bar{V}_{1,3}\\
\vdots & \vdots & \vdots \\
\bar{V}_{B,1} & \bar{V}_{B,2} & \bar{V}_{B,3}\end{bmatrix}
\]

where the $K=3$ metrics are the criteria (columns) and the rows are the mean score sample value $\bar{V}_{b,j}$ for each of the $B=15$ items (systems) to be ranked.  Using input matrix $M$, we calculate\footnote{Using the \texttt{pymcdm} module \cite{kizielewicz2023pymcdm}.} scores for each of the $B$ systems using the TOPSIS \citep{Hwang1981}, VIKOR \citep{duckstein1980multiobjective}, WSM and WPM \citep{fishburn1968sensitivity}, ERVD \citep{shyur2015multiple}, and SPOTIS \citep{dezert2020spotis} algorithms.  We then compare the ordering of these algorithms' system scores with those of the mean metric aggregate values $\bar{V}_{b,\textrm{agg}}$ for $b=1,\dots,B$ calculated by our framework in step~\ref{enum:agg_step}, for each dataset.

\begin{figure}[h!]
\centering
\includegraphics[width=0.91\columnwidth]{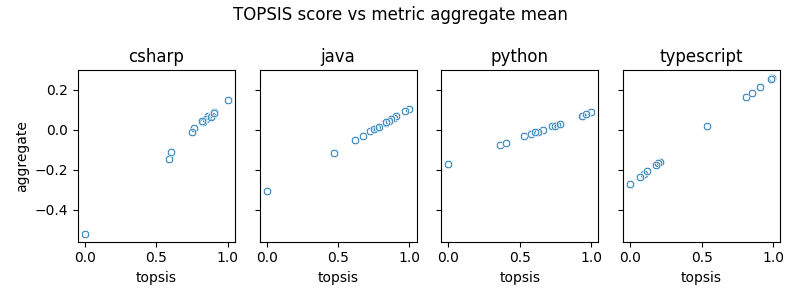}
\caption{Scatterplots of TOPSIS and mean metric aggregate scores by system, for each CrossCodeEval language dataset.}
\vspace{-0.1in}
\label{fig:topsis}
\end{figure}

The scatterplots of the TOPSIS vs our aggregate scores are shown in Figure~\ref{fig:topsis}; the orderings are identical except for C\#, where there is a minor difference of three consecutive system pairs switching order.  The results for the other MCDM algorithms were very similar, and the visual correlation between the scores was very strong throughout.  As discussed in Section~\ref{sec:introduction}, relying on a cross-metric aggregation for system evaluation, as we recommend, may obscure differences in metrics' measurements of systems that an MCDM technique like TOPSIS can exploit; this gap between aggregation and true MCDM may grow as number of criteria aggregated increases.  However, our framework has the advantage that it can quantify statistical significance of system comparisons and perform the type of dataset aggregation that algorithms like TOPSIS are not designed to do.  Nevertheless, the strong rank correspondence of our results with those of TOPSIS provides strong support for the soundness of our metric aggregation approach.

\section{Conclusion
\label{sec:conclusion}}

We present an open-source framework implementation to automate statistical testing comparisons of LLM-based systems by multiple evaluation criteria, and visualize the results.  The framework utilities also properly aggregate metric values and statistical tests across multiple benchmark datasets.  The statistically-grounded comparisons are designed to support user decisions of which LLM systems perform the best on benchmarks and metrics of interest.  We demonstrate the framework on the CrossCodeEval set of programming language benchmarks, on 15 state-of-the-art LLMs.
\section{Limitations
\label{sec:limitations}}
This work makes a valuable general contribution but is subject to several limitations.

Our framework was demonstrated on one benchmark, CrossCodeEval, albeit one that contains multiple component datasets.  Furthermore, the LLM systems considered were only openly available and not commercial.  The results were obtained from experiments from another paper.  However, the goal of the paper was to present a general framework to evaluate any metrics, systems, and datasets, rather than to demonstrate a specific predictive technique.  Therefore, this is not a crucial limitation since ultimately the quality of an evaluation depends on the choice of metrics and datasets which can effectively measure the system properties of interest.

Our analysis included a comparison with TOPSIS, a well-known MCDM method, as an illustration of the soundness of our approach to ranking systems on a comparable scenario (a single dataset with multiple metrics).  A more comprehensive comparison would involve other MCDM methods.  Furthermore, although our framework accepts unequal weights for metrics and datasets in aggregation, due to space constraints we did not evaluate how varying the weights affects our framework's analysis or impacts the comparison with TOPSIS when using the same weights.

\bibliography{main}



%
%





%

%

\onecolumn
\runningtitle{Supplementary materials}
\section{Auxiliary visualizations
\label{sec:auxiliary_visualizations}
}

\subsection{Grouping on auxiliary variables
\label{ssec:grouping_on_variables}
}

The analysis presented thus far has operated only on the values of the score samples $V_i$.  However, consider the setting where we have additional (i.e., `meta-data') variables $(z_1,\dots,z_k)$ in each dataset $D$.  For instance, continuing with the CrossCodeEval setting, say $z_1$ is some categorization of the type of code in input $x_i$ (e.g., hard/medium/easy task).  We may want to, say, conduct statistical comparisons on the systems separately at each combination of levels of one or more meta-features; perhaps two systems perform differently only on $z_1=`\textrm{hard}'$ instances but not on the others.  The method \texttt{signif\_pair\_diff\_bygroup}
conducts the grouping and performs the statistical tests, which may be used as input to the visualizations (Section~\ref{sec:visualizations}).

\subsection{Condition-and-compare system hyperparameters
\label{ssec:conditional_parameter_comparison}
}

This utility \texttt{compare\_system\_metrics\_conditioned\_on\_features} addresses a common scenario in hyperparameter optimization in testing of NLP or other ML-based systems.  Assume that we have a fixed pipeline where the component hyperparameters (see Section~\ref{sec:introduction}) can be denoted $\theta_1,\dots,\theta_k$.  For instance, $\theta_1$ is the LLM model, $\theta_2$ is the choice of prompt template, $\theta_3$ is the generation temperature, etc.  We are interested in one or more evaluation metrics $m_i$.  We have a a dataset $D$ of size $n$, and we want to compare the performance of the pipeline at the various values of hyperparameters $\theta_i$.

As before, a \textit{system} is defined by each unique combination values of the $k$ hyperparameters.  Say that we create a static set of $B\leq n$ such systems, such as by a factorial experiment or independent random sampling of values of each $\theta_i$.  Now, each input $x_i\in D$ is run on a randomly selected system out of the $B$.  We now have a \textit{paired} experimental setup within each system, where $L_j=(V_{1,j},\dots,V_{B,j})$ for each metric $j$, and each score sample $V_{b,j}$ is measured only on the same $n_{b,j}=|V_{b,j}|$ observations in $D$ that were run on system $b$.  However, across systems $b$ there is not pairing since the observation sets for each system $b$ are mutually exclusive.  Since it is possible that $n/B\geq 1$ is small, i.e., each unique system is run on only one or a small number of observations, but $B$ can be large, a full pairwise comparison of the $B$ systems can be computationally difficult due to the large size of $B$, and each statistical test will be be weak due to the small score sample sizes.  

Thus, we want a more restricted analysis to help understand the effects of values of the different hyperparameters $\theta_i$.  Assume for simplicity that the hyperparameter space for each $\theta_i$ for creating the $B$ systems is a small discrete set of values, not a continuous space.  If more than one metric is of interest, we create a new aggregate metric (Section~\ref{ssec:sample_aggregation}) from the full set of paired observations $D$, which is the new evaluation metric.  Let us specify sets $\Theta_{\textrm{compare}}$ of at least size 1, and possibly empty $\Theta_{\textrm{condition}}$, which are mutually subsets of the hyperparameters $\theta$.  If  |$\Theta_{\textrm{condition}}|=0$, the full $D$ is one group comprising all $B$ systems.  Otherwise, the $B$ systems are partitioned conditioning on the unique observed combinations of values of hyperparameters $\Theta_{\textrm{condition}}$. Then, in each such partition subset, the system evaluation metric sample scores are pooled according to the unique combination of hyperparameters $\Theta_{\textrm{compare}}$ observed. The metric distributions per unique comparison value are then either plotted (Section~\ref{sec:exploratory_visualization}) or statistically compared via the post-op procedure (Section~\ref{ssec:post_op_plotting_procedure}) with `first vs the rest' iteration.  Systems in the visualizations can further be colored according to unique combinations of a further set of hyperparameters $\Theta_{\textrm{color}}\subseteq\Theta_{\textrm{compare}}$.

\begin{figure}
\centering
\includegraphics[width=0.48\columnwidth]{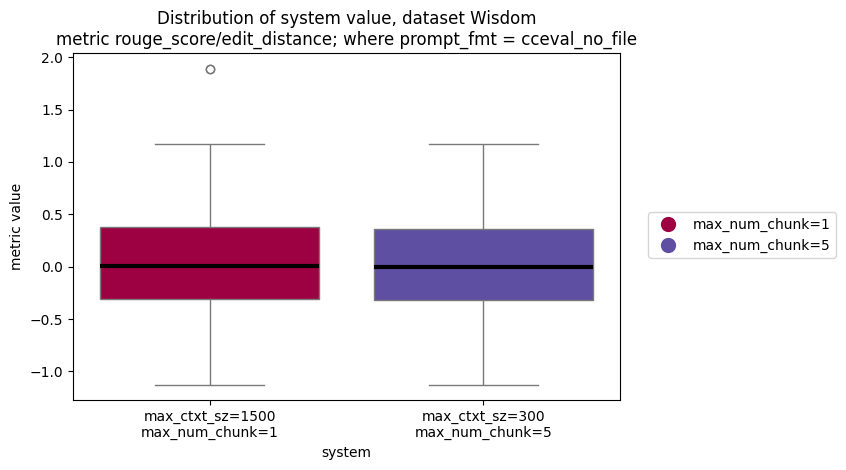}
\includegraphics[width=0.48\columnwidth]{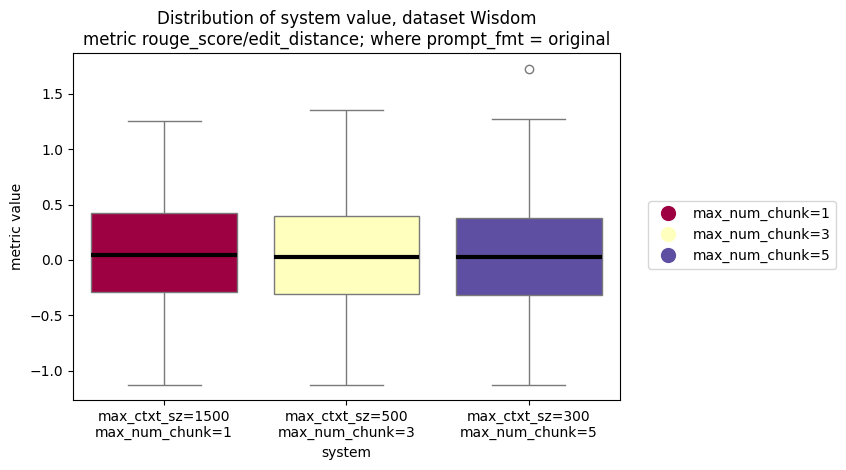}
\caption{Condition-and-compare analysis for an example dataset.
}
\vspace{-0.1in}
\label{fig:parameter_comparison}
\end{figure}

Figure~\ref{fig:parameter_comparison} shows an example of the two plots produced on an example dataset with evaluation metrics $m_1$=Rouge-score and $m_2$=Edit distance, and $B=8$ different systems defined by hyperparameters $\theta_1=\textrm{max\_ctxt\_sz}$, $\theta_2=\textrm{max\_num\_chunk}$, $\theta_3=\textrm{snip\_thresh}$, $\theta_4=\textrm{prompt\_fmt}$.  The analysis is done by conditioning on 
$\Theta_{\textrm{condition}}=(\theta_4)$, at each group comparing systems based on $\Theta_{\textrm{compare}}=(\theta_1,\theta_2)$ and coloring based on $\Theta_{\textrm{color}}=(\theta_2)$, with an aggregation of the metrics.  The two panels show boxplots of the aggregate evaluation metric at each observed value $(\textrm{cceval\_no\_file}, \textrm{original})$ of the conditioning variable $\theta_4$.  The conclusion is that, visually, for each prompt format ($\theta_4$), that the choice of values of hyperparameters $\theta_1,\theta_2$ does not seem to affect the evaluation metric value.  It is possible that when removing the conditioning on $\theta_4$ there would be an effect across $\theta_1,\theta_2$.

\end{document}